\documentclass[preprintnumbers, floatfix,letterpaper,aps,prd,epsfig,nofootinbib,
twocolumn
]{revtex4-2}

\usepackage{graphicx}% Include figure files
\usepackage{dcolumn}% Align table columns on decimal point
\usepackage[colorlinks,linkcolor=blue,citecolor=blue,urlcolor=blue]{hyperref}
\usepackage{bm,graphicx,dcolumn,epstopdf,epsf, latexsym,mathbbol, amssymb,amsmath,color, slashed, mathrsfs,mathcomp,simplewick}
\usepackage[center]{subfigure}
\usepackage{multirow}
\usepackage{makecell}

 \newcommand{\bq}{\begin{equation}}
 \newcommand{\eq}{\end{equation}}
 \newcommand{\bqn}{\begin{eqnarray}}
 \newcommand{\eqn}{\end{eqnarray}}
 \newcommand{\nb}{\nonumber}
 \newcommand{\lb}{\label}

\begin{document}

\title{Improved constraints on higher-derivative effective field theory extensions of general relativity with GW230627\_015337}

\author{Tao Zhu}
\email{zhut05@zjut.edu.cn}
\affiliation{
Institute for theoretical physics \& cosmology, Zhejiang University of Technology, Hangzhou, 310032, China}
\affiliation{United Center for Gravitational Wave Physics, Zhejiang University of Technology, Hangzhou, 310032, China}

\date{\today}

\begin{abstract}

Gravitational-wave (GW) observations of binary black holes provide a powerful probe of general relativity (GR) in the strong-field regime. In effective field theory (EFT) extensions of GR, higher-curvature operators modify the binary dynamics and introduce corrections to the inspiral phase. In this work, we consider two parity-even cubic-curvature couplings and one quartic-curvature coupling, whose leading phase corrections enter at 5PN, 6PN, and 8PN order, respectively. We incorporate these corrections into the individual multipoles of the \texttt{SEOBNRv5HM\_ROM} waveform using the flexible theory-independent framework, allowing higher-order modes and the full inspiral-merger-ringdown signal to be used in the analysis. We first perform a Fisher-matrix analysis of 92 binary-black-hole events selected from a parent sample of 273 events in the public LIGO-Virgo-KAGRA catalogs. GW230627\_015337 provides the tightest forecast constraints on all three EFT couplings and is therefore selected for a full Bayesian analysis. The posterior distributions are consistent with GR, with all three couplings consistent with zero at 90\% credibility. The corresponding one-sided 90\% upper limits on the characteristic length scales are \(17.1\), \(11.1\), and \(18.3\,\mathrm{km}\). The log Bayes factors of the three EFT models relative to GR range from \(-5.49\) to \(-2.51\), favoring GR under the adopted priors. These bounds probe higher-curvature length scales at the level of \(10\)--\(20\,\mathrm{km}\) and are numerically tighter than representative previous inspiral and ringdown constraints, although differences in the models and analysis assumptions should be taken into account.

\end{abstract}

\maketitle

\section{Introduction}\label{sec:intro}
\renewcommand{\theequation}{1.\arabic{equation}} \setcounter{equation}{0}

The detection of gravitational waves (GWs) from coalescing compact binaries by the LIGO-Virgo-KAGRA (LVK) Collaboration \cite{LIGOScientific:2016aoc,LIGOScientific:2016vlm,LIGOScientific:2016emj,LIGOScientific:2017vwq,LIGOScientific:2018mvr,LIGOScientific:2020ibl,KAGRA:2021vkt,LIGOScientific:2025slb,LIGOScientific:2026wfs} has opened a new observational window for testing gravity in the strong-field and highly dynamical regime. The observed signals carry information about the binary dynamics throughout the inspiral, merger, and ringdown stages. In particular, the large number of orbital cycles accumulated during the inspiral makes the GW phase especially sensitive to small modifications of the binary dynamics. Compact-binary GW signals therefore provide a powerful probe of general relativity (GR) and possible deviations from it. So far, tests based on the published GW catalogs have found no statistically significant evidence for departures from GR \cite{LIGOScientific:2019fpa,LIGOScientific:2020tif,LIGOScientific:2021sio,LIGOScientific:2026fcf,LIGOScientific:2026oim}. With the continuing improvement in detector sensitivity and the growing number of observed events, GW observations are expected to place increasingly stringent constraints on possible extensions of GR.

Tests of GR with GWs can be broadly divided into theory-agnostic and theory-specific approaches. In theory-agnostic tests, possible deviations from GR are introduced directly at the waveform level through phenomenological parameters, without assuming a particular gravitational theory. Examples include the parametrized post-Einsteinian framework \cite{Tahura:2018zuq,Yunes:2009ke}, parametrized modifications to GW propagation associated with symmetry breaking \cite{Zhu:2023rrx,Wang:2025fhw}, and inspiral tests in which selected post-Newtonian (PN) phase coefficients are allowed to deviate from their GR predictions. The latter have been implemented, for example, in the TIGER pipeline \cite{Agathos:2013upa,Li:2011cg,Meidam:2017dgf,Roy:2025gzv} and in the flexible theory-independent (FTI) framework \cite{Mehta:2022pcn}. These methods are well suited for searching for generic deviations from GR. However, the phenomenological parameters introduced in such tests do not always admit a direct interpretation in terms of the fundamental parameters of a specific theory. Theory-specific tests follow a complementary strategy. Starting from a given gravitational action, one derives the modifications to the binary dynamics and GW waveform and then uses GW observations to constrain the corresponding coupling parameters. In this work, we follow this approach and consider higher-curvature corrections to GR within the framework of effective field theory (EFT).

EFT provides a systematic description of corrections to GR below a cutoff scale \(\Lambda\) \cite{Donoghue:1995cz,Donoghue:1994dn,Endlich:2017tqa,Cardoso:2018ptl,Sennett:2019bpc,deRham:2020ejn}. For four-dimensional vacuum binary-black-hole (BBH) systems, curvature-squared terms are either topological or can be removed by field redefinitions at the order relevant here \cite{Sennett:2019bpc,Liu:2023spp}. We therefore focus on cubic- and quartic-curvature interactions. In the cubic sector, we consider a parity-even model with two independent coupling parameters, whose effects on the binary binding energy, GW flux, and inspiral phase have been derived in Refs.~\cite{AccettulliHuber:2020dal,Liu:2024atc,Liu:2023spp}. We also consider the parity-even quartic operator \(\mathcal{C}^{2}\) studied in Ref.~\cite{Sennett:2019bpc}. This represents a particular truncation of the quartic-curvature sector rather than the most general EFT operator basis. The leading inspiral-phase corrections enter at 5PN and 6PN order for the two cubic couplings and at 8PN order for the quartic coupling. In the present analysis, we include these EFT corrections to the inspiral phase and neglect corrections to the waveform amplitudes and spin-dependent terms.

Higher-curvature EFT extensions of GR have previously been constrained using both inspiral and ringdown signals. Ringdown analyses probe modifications to the quasinormal-mode spectrum of the remnant black hole. Using GW150914 and GW200129, Silva et al. obtained 90\% upper limits of \(\ell_{\rm qEFT}\leq 51.3\) km and \(\ell_{\rm cEFT}\leq 38.2\) km for the quartic and cubic models considered in that analysis, respectively \cite{Silva:2022srr}, with the quartic bound depending on the adopted EFT-validity condition. Inspiral analyses instead probe changes in the binary dynamics through their accumulated effects on the GW phase. An analysis of GW151226 and GW170608 found that quartic-EFT length scales around \(d_\Lambda\equiv1/\Lambda\sim150\) km are disfavored \cite{Sennett:2019bpc}, while the cubic sector has also been constrained using GW170608 \cite{Liu:2024atc}. Since these analyses probe different parts of the signal and rely on different waveform models, frequency ranges, priors, and validity assumptions, their constraints are complementary.

For inspiral-based EFT tests, an important issue is how to incorporate the predicted phase corrections into modern waveform models while retaining as much information as possible from the observed signal. Many PN-based tests use inspiral-only waveforms or focus mainly on the dominant \((2,2)\) mode. However, higher-order multipoles can provide additional information, especially for binaries with unequal masses or non-face-on orientations. Moreover, although the EFT corrections considered here are derived only for the inspiral, the merger and ringdown portions of the signal still help constrain the binary masses, spins, distance, and orientation, and can therefore reduce degeneracies with beyond-GR effects.

The FTI framework provides a suitable way to incorporate these features \cite{Mehta:2022pcn}. It allows parametrized phase corrections to be introduced into the individual multipoles of a complete GR inspiral-merger-ringdown waveform. The modifications are applied in the inspiral and smoothly tapered toward its end, while the calibrated GR merger-ringdown waveform is retained. In this way, higher-order modes can be included and the full observed signal can be used in parameter estimation even when the beyond-GR corrections are known only in the inspiral regime. This is particularly relevant for the high-PN EFT corrections considered here, which become more important toward the late inspiral and can be correlated with the binary masses and spins. We therefore use the FTI framework to introduce the 5PN, 6PN, and 8PN EFT phase corrections into the individual multipoles of the \texttt{SEOBNRv5HM\_ROM} waveform, with the PN coefficients expressed directly in terms of the EFT coupling parameters.

A further question is which BBH events are expected to provide the strongest constraints on these EFT corrections. A full Bayesian analysis of every detected BBH event would be computationally expensive, especially when higher-order modes and a complete inspiral-merger-ringdown waveform are included. It is therefore useful to first identify the most promising events with a computationally cheaper method and then perform a full Bayesian analysis for the best candidates. However, the network signal-to-noise ratio (SNR) alone is not sufficient for such a selection. The constraining power also depends on the source mass, the amount of inspiral signal in band, the frequency dependence of the EFT correction, parameter correlations, detector responses, and noise power spectral densities. In addition, converting a dimensionless EFT coupling to a characteristic length scale introduces an extra dependence on the source mass. Therefore, a louder event does not necessarily provide a tighter constraint on the EFT length scale. This motivates our use of the Fisher-matrix analysis to rank the observed BBH events and select the most promising target for the subsequent Bayesian inference.

In this work, we proceed in three steps. First, we construct EFT-modified waveforms by introducing the cubic- and quartic-EFT phase corrections at 5PN, 6PN, and 8PN order into the individual multipoles of \texttt{SEOBNRv5HM\_ROM} using the FTI prescription \cite{Pompili:2023tna,Mehta:2022pcn}. Second, we compile a parent sample of 273 BBH events from the public GWTC catalogs and select 92 events with sufficient inspiral information for a Fisher-matrix analysis \cite{LIGOScientific:2025slb,LIGOScientific:2026wfs}. We use the Fisher forecasts to rank these events according to their expected constraining power on the EFT parameters and identify GW230627\_015337 as the most promising target in our sample. Finally, we perform a full Bayesian analysis of this event, obtain posterior distributions for the EFT coupling parameters and the corresponding characteristic length scales, and compare the resulting constraints with the Fisher forecasts and with previous inspiral- and ringdown-based bounds.

The remainder of this paper is organized as follows. In Sec.~\ref{sec:2}, we introduce the higher-derivative EFT extension of GR, derive the corresponding modifications to the GW phase, and describe how these corrections are incorporated into the GR waveform within the FTI framework. In Sec.~\ref{sec:3}, we describe the Fisher-matrix formalism and event-selection criteria, present the dependence of the forecast constraints on the binary mass, and rank the selected BBH events according to their expected constraining power. In Sec.~\ref{sec:4}, we describe the Bayesian inference method and the setup used in our analysis. The main results are presented in Sec.~\ref{sec:5}, where we also compare our constraints with previous results. Finally, we summarize our main findings in Sec.~\ref{sec:6}.

Throughout this paper, we adopt the metric signature \((- , + , + , +)\). Greek indices \((\mu,\nu,\ldots)\) denote spacetime coordinates and run over \(0,1,2,3\), while Latin indices \((i,j,k,\ldots)\) denote spatial coordinates and run over \(1,2,3\). Unless otherwise stated, we keep the factors of \(G\) and \(c\) explicit.

\section{The effective field theory extension to general relativity and the modified waveforms}\label{sec:2}
\renewcommand{\theequation}{2.\arabic{equation}} \setcounter{equation}{0}

\subsection{The higher-derivative effective field theory extension of GR}

As discussed in the Introduction, we describe higher-curvature corrections to GR through a low-energy derivative expansion of the gravitational action. In this section, we specify the operators retained in our analysis and define the corresponding coupling conventions. We follow the two-body effective description and coupling conventions adopted in Refs.~\cite{AccettulliHuber:2020dal,Liu:2024atc}. These conventions will also be used below to construct the waveform corrections and to relate the inferred couplings to characteristic length scales.

At four-derivative order, the parity-even pure-gravity action contains three curvature-squared invariants, which may be chosen as
\(R^2\), \(R_{\mu\nu}R^{\mu\nu}\), and
\(R_{\mu\nu\rho\sigma}R^{\mu\nu\rho\sigma}\).
In four dimensions, the Riemann-squared term can be expressed, up to the Gauss--Bonnet density, in terms of the Ricci scalar and Ricci tensor. The latter can be removed perturbatively by metric field redefinitions using the leading-order vacuum equations of motion, while the Gauss--Bonnet density with a constant coefficient is topological and does not modify the local field equations \cite{Sennett:2019bpc}. The parity-odd curvature-squared invariant is the Chern--Pontryagin density,
\begin{align}
\widetilde{\mathcal C}
&\equiv
R_{\mu\nu\rho\sigma}\widetilde R^{\mu\nu\rho\sigma},
\\
\widetilde R_{\mu\nu\rho\sigma}
&\equiv
\frac{1}{2}\epsilon_{\rho\sigma}{}^{\alpha\beta}
R_{\mu\nu\alpha\beta},
\end{align}
which is also a total derivative for a constant coefficient. We therefore do not include independent curvature-squared operators in the vacuum gravitational action considered here. We note, however, that field redefinitions in the bulk action can induce or shift operators in the effective source or worldline action. The reduction of the vacuum operator basis should therefore be distinguished from the operator content of the complete compact-binary EFT \cite{AccettulliHuber:2020dal,Liu:2024atc}.

We next consider the six-derivative cubic-curvature sector. Following Refs.~\cite{AccettulliHuber:2020dal,Liu:2024atc}, we introduce
\begin{align}
I_1 &=
R_{\mu\nu}{}^{\alpha\beta}
R_{\alpha\beta}{}^{\gamma\sigma}
R_{\gamma\sigma}{}^{\mu\nu},
\\
I_2 &=
R_{\mu}{}^{\alpha}{}_{\nu}{}^{\beta}
R_{\alpha}{}^{\gamma}{}_{\beta}{}^{\sigma}
R_{\gamma}{}^{\mu}{}_{\sigma}{}^{\nu},
\end{align}
and define the combination
\begin{align}
\mathcal G_3\equiv I_1-2I_2.
\end{align}
The combination \(\mathcal G_3\) is related to the purely Riemann-cubic part of the six-dimensional Euler density. Although the corresponding structures are not independent in the four-dimensional vacuum on-shell basis, both couplings enter the two-body effective description adopted in Refs.~\cite{AccettulliHuber:2020dal,Liu:2024atc}. We therefore follow these references and write
\begin{align}
{\cal L}_{\rm cubic}
=
\alpha_1 I_1
+\frac{\alpha_2}{2}\mathcal G_3 .
\label{eq:Lcubic}
\end{align}
Parity-odd cubic operators can also be constructed by dualizing one of the Riemann tensors. These operators are not included in the present analysis \cite{Brandhuber:2024qdn}.

At eight-derivative order, the action can contain quartic-curvature interactions. We define
\begin{align}
\mathcal C
&\equiv
R_{\alpha\beta\gamma\delta}
R^{\alpha\beta\gamma\delta},
&
\widetilde{\mathcal C}
&\equiv
R_{\alpha\beta\gamma\delta}
\widetilde R^{\alpha\beta\gamma\delta}.
\end{align}
These quantities generate the quartic operators
\(\mathcal C^2\),
\(\widetilde{\mathcal C}^{\,2}\), and
\(\mathcal C\widetilde{\mathcal C}\).
The first two are parity even, whereas
\(\mathcal C\widetilde{\mathcal C}\) is parity odd \cite{Sennett:2019bpc}. Following Ref.~\cite{Sennett:2019bpc}, we retain only the parity-even operator \(\mathcal C^2\),
\begin{align}
{\cal L}_{\rm quartic}
=
\alpha_3\mathcal C^2.
\label{eq:Lquartic}
\end{align}
This choice represents a truncation of the quartic-curvature sector rather than the most general EFT operator basis.

Combining the retained cubic and quartic interactions with the Einstein--Hilbert term, the gravitational action considered in this work is
\begin{eqnarray}
S_{\rm g}
&=&
\frac{1}{16\pi G}
\int d^4x\,\sqrt{-g}\,
\bigg[
R
+\frac{\alpha_1}{\Lambda^4} I_1
+\frac{\alpha_2}{2\Lambda^4}\mathcal G_3
+\frac{\alpha_3}{\Lambda^6}\mathcal C^2
\bigg]. \nonumber\\
\label{eq:action}
\end{eqnarray}
Here, \(\Lambda\) is the EFT cutoff scale, while
\(\alpha_1\), \(\alpha_2\), and \(\alpha_3\) are dimensionless coupling coefficients. Equation~\eqref{eq:action} defines the truncated higher-curvature model considered throughout this work. Setting
\(\alpha_1=\alpha_2=\alpha_3=0\)
recovers the Einstein--Hilbert action.

For quasicircular binary inspirals, these higher-curvature interactions modify the conservative dynamics and GW energy flux, and consequently lead to corrections to the waveform phase. In the nonspinning limit, the leading phase corrections associated with \(\mathcal G_3\), \(I_1\), and \(\mathcal C^2\) enter at 5PN, 6PN, and 8PN order, respectively \cite{AccettulliHuber:2020dal,Liu:2024atc,Sennett:2019bpc}. The explicit forms of these phase corrections are given in the next subsection.

\subsection{The EFT phase corrections}

We introduce the dimensionless PN velocity parameter associated with the dominant \((2,2)\) mode,
\bqn
v \equiv \left(\pi M f\right)^{1/3},
\eqn
where, in this subsection, we use units with \(G=c=1\). Here \(M=m_1+m_2\) is the detector-frame total mass and \(f\) is the observed GW frequency of the \((2,2)\) mode. We also define the symmetric mass ratio as
\(\eta\equiv m_1m_2/M^2\).
Since \(M=(1+z)M_{\rm src}\) and \(f=f_{\rm src}/(1+z)\), the combination \(Mf=M_{\rm src}f_{\rm src}\) is independent of the cosmological redshift. The EFT phase corrections used below are those derived for quasicircular, nonspinning binaries.

At the PN orders considered here, the EFT corrections can be written as shifts to the PN coefficients of the \((2,2)\)-mode phase,
\bqn
\delta\Psi^{\rm EFT}(f)
=
\frac{3}{128\eta}
\sum_{n\in\{10,12,16\}}
\delta\psi_n\,v^{\,n-5}.
\lb{eq:psi_gen}
\eqn
Here \(n=2k\), where \(k\) denotes the PN order relative to the leading Newtonian phase. The corrections proportional to
\(\delta\psi_{10}\),
\(\delta\psi_{12}\), and
\(\delta\psi_{16}\)
therefore enter at 5PN, 6PN, and 8PN order, respectively. Their explicit expressions are
\bqn
\delta\psi_{10}
&=&
-1872\,\beta_2,
\nb\\
\delta\psi_{12}
&=&
\frac{
549360\,\beta_1
-45(43683+12908\eta)\beta_2
}{294},
\nb\\
\delta\psi_{16}
&=&
\frac{234240-522240\,\eta}{11}\,\beta_3 .
\lb{eq:psi_shifts}
\eqn
The corresponding dimensionless coupling combinations are defined by
\bqn
\beta_1
&\equiv&
\alpha_1
\left(\frac{\ell_\Lambda}{r_g}\right)^4,
\nb\\
\beta_2
&\equiv&
\alpha_2
\left(\frac{\ell_\Lambda}{r_g}\right)^4,
\nb\\
\beta_3
&\equiv&
\alpha_3
\left(\frac{\ell_\Lambda}{r_g}\right)^6 ,
\lb{eq:beta_def}
\eqn
where \(\ell_\Lambda\equiv 1/\Lambda\) is the EFT length scale and
\(r_g=M_{\rm src}\) is the gravitational radius of the binary in the source frame in units of \(G=c=1\). When units are restored,
\(r_g=GM_{\rm src}/c^2\).

The cubic-sector coefficients in Eq.~\eqref{eq:psi_shifts} are taken from Eqs.~(25)--(26) of Ref.~\cite{Liu:2024atc}, which revises the earlier calculation of Ref.~\cite{AccettulliHuber:2020dal}, while the 8PN term is the leading phase correction associated with the quartic operator retained here \cite{Sennett:2019bpc}. 
%An important feature of the cubic sector is that the PN corrections are not independent. In particular, \(\beta_2\) contributes to both the 5PN and 6PN terms, whereas \(\beta_1\) first enters at 6PN. Thus, when a given EFT coupling is varied, the corresponding corrections at different PN orders are varied consistently according to Eq.~\eqref{eq:psi_shifts}. 
Higher-PN contributions beyond those shown above are not included in the present analysis. The GR limit is recovered for
\(\beta_1=\beta_2=\beta_3=0\).

We perform three separate single-coupling analyses. In each case, one of the parameters \(\beta_i\) is allowed to vary while the other two are fixed to zero. We also consider GR as the reference model, for which all three EFT couplings vanish. The constraints obtained below should therefore be understood as bounds on individual one-dimensional directions in the EFT coupling space, rather than joint constraints obtained by varying all three couplings simultaneously.

Although the waveform model is parametrized in terms of the dimensionless quantities \(\beta_i\), it is more convenient to compare constraints from different sources and previous studies in terms of characteristic length scales. We define
\bqn
\ell_1
&\equiv&
|\alpha_1|^{1/4}\ell_\Lambda
=
|\beta_1|^{1/4}r_g,
\nb\\
\ell_2
&\equiv&
|\alpha_2|^{1/4}\ell_\Lambda
=
|\beta_2|^{1/4}r_g,
\nb\\
\ell_3
&\equiv&
|\alpha_3|^{1/6}\ell_\Lambda
=
|\beta_3|^{1/6}r_g .
\lb{eq:ell_def}
\eqn
The quantities \(\ell_i\) on the left-hand side characterize the underlying EFT and are independent of the individual source, while the expressions on the right-hand side provide the mapping from the event-level constraints on \(\beta_i\) to the corresponding EFT length scales. The absolute values imply that \(\ell_i\) characterize only the magnitudes of the couplings. The signs of the couplings remain relevant to the waveform phase and are retained in the inference on \(\beta_i\).

\subsection{Implementation of the EFT phase corrections in the FTI framework}
\label{sec:fti}

The EFT phase corrections derived in the previous subsection describe the inspiral of a quasicircular, nonspinning binary, whereas the waveform model used in our data analysis also contains higher-order modes and the merger-ringdown signal. To combine these ingredients, we use the FTI prescription of Ref.~\cite{Mehta:2022pcn}. The FTI framework introduces inspiral phase corrections into individual frequency-domain modes and smoothly tapers these corrections toward the end of the inspiral, while retaining the underlying GR waveform in the post-inspiral regime. Although the FTI framework is theory independent in its general form, here the phase corrections are fixed by the EFT model described in the previous subsections.

For a positive-\(m\) harmonic, the frequency-domain phase can be written as
\bqn
\Psi_{\ell m}(f)
&=&
2\pi f t_c-m\phi_c-\frac{\pi}{4}
\nb\\
&&
+\frac{3}{128\eta}\frac{m}{2}v_m^{-5}
\sum_n
\Big[
\left(\varphi_n+\delta\varphi_n\right)  \nonumber \\
&&\;\;\;\;\;\;\;\;\;\;\;\;\;\;
+\left(\varphi_{n\ell}+\delta\varphi_{n\ell}\right)
\ln v_m
\Big]v_m^n ,
\lb{eq:fti_phase}
\eqn
where \(t_c\) and \(\phi_c\) denote the coalescence time and orbital reference phase, respectively. The mode-dependent PN velocity is defined by
\bqn
v_m
\equiv
\left(\frac{2\pi M f}{m}\right)^{1/3},
\eqn
where \(M\) is the detector-frame total mass and \(f\) is the observed GW frequency of the corresponding mode. For the dominant \((2,2)\) mode, this reduces to
\(v_2=(\pi Mf)^{1/3}=v\).
The quantities \(\varphi_n\) and \(\varphi_{n\ell}\) denote the nonlogarithmic and logarithmic GR phase coefficients, while
\(\delta\varphi_n\) and
\(\delta\varphi_{n\ell}\) denote their corresponding shifts. A term proportional to \(v_m^{n-5}\) enters at \(n/2\) PN order relative to the leading Newtonian phase.

In a generic FTI analysis, the phase-coefficient deviations can be treated as independent parameters. In the present theory-specific implementation, we instead identify the EFT corrections in Eqs.~(\ref{eq:psi_gen})--(\ref{eq:psi_shifts}) with the corresponding FTI coefficient shifts,
\bqn
\delta\varphi_n
&=&
\delta\psi_n,
\qquad
n\in\{10,12,16\},
\nb\\
\delta\varphi_n
&=&
0,
\qquad
n\notin\{10,12,16\},
\nb\\
\delta\varphi_{n\ell}
&=&
0 .
\eqn
Thus, no additional logarithmic corrections are introduced. The EFT relations among different PN orders are retained. In particular, \(\beta_2\) determines both the 5PN and 6PN corrections according to Eq.~(\ref{eq:psi_shifts}), rather than allowing the two coefficients to vary independently.

We apply the same EFT phase prescription to each retained waveform mode using the \(m/2\) scaling in Eq.~(\ref{eq:fti_phase}). This should be understood as the mode-by-mode FTI implementation of the inspiral correction, rather than as an independent EFT calculation of the phase correction for each higher-order mode.

We implement the tapering procedure following Ref.~\cite{Mehta:2022pcn}, using the LALSuite routine
\texttt{SimInspiralTestingGRCorrections} \cite{LALSuite} and the mode-by-mode implementation in \texttt{bilby\_tgr} \cite{bilbyTGR}. The baseline GR frequency-domain waveform is generated with
\texttt{SEOBNRv5HM\_ROM} \cite{Pompili:2023tna}. We retain the seven positive-\(m\) modes
\((2,2)\), \((2,1)\), \((3,3)\), \((3,2)\), \((4,4)\), \((4,3)\), and \((5,5)\).
The phase correction is applied separately to each mode before the plus and cross polarizations are reconstructed.

For each mode, the untapered EFT phase correction is
\bqn
\delta\Psi_{\ell m}(f)
=
\frac{3}{128\eta}\frac{m}{2}
\sum_{n\in\{10,12,16\}}
\delta\psi_n\,v_m^{\,n-5}.
\lb{eq:untapered_phase}
\eqn
Equation~(\ref{eq:untapered_phase}) follows the phase convention used in the analytic expressions of the previous subsection. The frequency-domain phase convention used by LALSuite has the opposite sign,
\(\Psi_{\rm LAL}=-\Psi_{\rm lit}\).
The corresponding correction to the LALSuite waveform is therefore
\(-\delta\Psi_{\ell m}\). We account for this sign difference when constructing the tapered correction below.

Following the FTI prescription, we taper the second frequency derivative of the phase correction using the Fermi--Dirac window
\bqn
W_{\ell m}(f)
=
\frac{1}{
1+\exp\left[(v_m-v_{\rm w})/\sigma_{\rm w}\right]
},
\label{eq:window}
\eqn
where
\bqn
v_m
=
\left(\frac{2\pi M f}{m}\right)^{1/3},
\qquad
v_{\rm w}
=
\left(\pi M F f_{22}^{\rm peak}\right)^{1/3}.
\eqn
Here \(f_{22}^{\rm peak}\) is the observed peak frequency of the GR \((2,2)\) mode, evaluated using the fitting formula of Ref.~\cite{Bohe:2016gbl}, and \(F\) controls the taper location. We adopt \(F=1\), following Ref.~\cite{LIGOScientific:2026oim}. Since the same value of \(v_{\rm w}\) is used for all modes, the taper corresponds to the same orbital frequency, while its Fourier-domain frequency depends on the harmonic index \(m\). In particular, the center of the window, defined by \(W_{\ell m}=1/2\), occurs at
\bqn
f=\frac{m}{2}F f_{22}^{\rm peak}.
\eqn
Thus, different harmonics are tapered at different GW frequencies, but at the same underlying orbital frequency.

%Following the FTI prescription, we taper the second frequency derivative of the phase correction. For each mode, we introduce the Fermi--Dirac window
%\bqn
%W_{\ell m}(f)
%=
%\frac{1}{
%1+\exp\left[(v_m-v_{\rm w})/\sigma_{\rm w}\right]
%}.
%\lb{eq:window}
%\eqn
%The center of the transition is specified by
%\bqn
%v_{\rm w}
%=
%\left(\pi M F f_{22}^{\rm peak}\right)^{1/3},
%\eqn
%where \(f_{22}^{\rm peak}\) is the observed peak frequency of the GR
%\((2,2)\) mode, evaluated using the fitting formula of
%Ref.~\cite{Bohe:2016gbl}. Since the taper is defined at a fixed orbital velocity, the corresponding center frequency of the \(m\)-th mode is
%\bqn
%f_{{\rm w},m}
%=
%\frac{m}{2}F f_{22}^{\rm peak}.
%\eqn
%The dimensionless parameter \(F\) controls the location of the transition relative to the \((2,2)\)-mode peak. We adopt \(F=1\), following Ref.~\cite{LIGOScientific:2026oim}.

The width of the transition is chosen according to the FTI prescription,
\bqn
\sigma_{\rm w}
=
\frac{128\eta}{3}\,
\pi v_{\rm w}^{6}
\frac{N_{\rm cyc}}{50},
\eqn
where \(N_{\rm cyc}\) controls the width of the transition in GW cycles. We use \(N_{\rm cyc}=1\), as in Ref.~\cite{Mehta:2022pcn}. The window approaches unity well below the transition and zero above it, with
\(W_{\ell m}=1/2\) at the center. 
%We emphasize that this taper specifies how the inspiral correction is connected to the GR post-inspiral waveform and does not by itself define the regime of validity of the EFT.

Taking into account the phase convention described above, we define the tapered second derivative as
\bqn
D_{\ell m}(f)
\equiv
-
W_{\ell m}(f)
\frac{d^2\delta\Psi_{\ell m}(f)}{df^2}.
\lb{eq:Dlm}
\eqn
The phase correction applied to the waveform is then obtained by integrating \(D_{\ell m}\) twice,
\bqn
Q_{\ell m}(f)
=
\int_{f_{{\rm ref},m}}^f dx
\int_{f_{{\rm a},m}}^x dy\,
D_{\ell m}(y),
\lb{eq:Qlm}
\eqn
where
\bqn
f_{{\rm ref},m}
=
\frac{m}{2}f_{\rm ref}
\eqn
is the mode-dependent reference frequency. In our implementation, the second integration anchor is chosen as
\bqn
f_{{\rm a},m}=f_{22}^{\rm peak}.
\eqn
The corresponding integration conditions are
\bqn
Q_{\ell m}(f_{{\rm ref},m})
&=&0,
\nb\\
Q'_{\ell m}(f_{{\rm a},m})
&=&0.
\lb{eq:bc}
\eqn
The first condition fixes the phase correction at the reference frequency, while the second fixes the linear-in-frequency integration constant according to the adopted FTI implementation. Since the two conditions are imposed at different frequencies,
\(Q'_{\ell m}(f_{{\rm ref},m})\) does not in general vanish.

The corrected waveform mode is finally given by
\bqn
\widetilde h_{\ell m}(f)
\longrightarrow
\widetilde h_{\ell m}(f)
\exp\left[iQ_{\ell m}(f)\right].
\lb{eq:corrected_mode}
\eqn
We set the correction to zero below the common analysis frequency
\(f_{\rm min}=20\,{\rm Hz}\), while retaining the native frequency support of each mode. The corrected modes are then combined to construct the plus and cross polarizations. In the present model, only the waveform phase is modified; EFT corrections to the mode amplitudes are not included.

\begin{figure*}[tbp]
\centering
\includegraphics[width=0.95\textwidth]{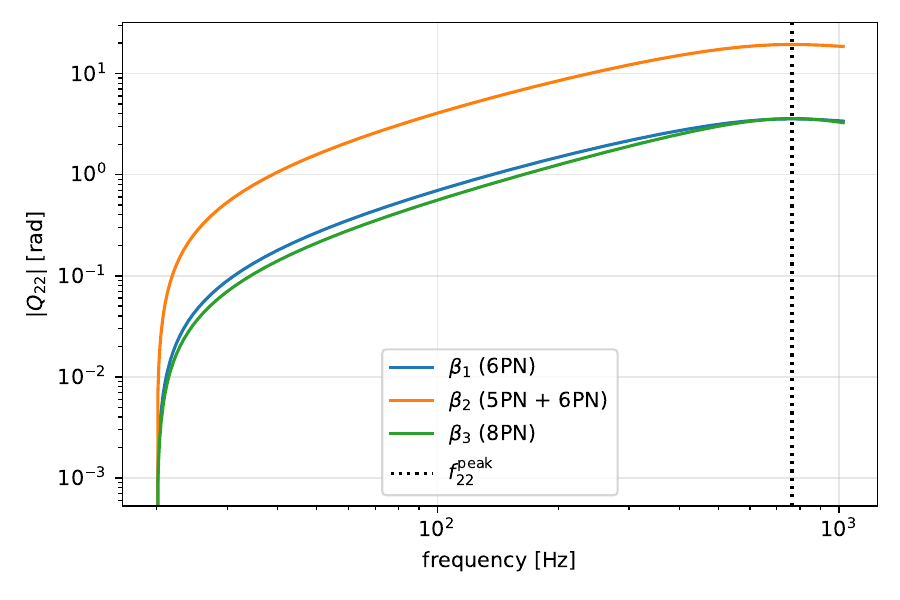}
\caption{\label{fig:dephasing}
Magnitude of the tapered phase correction, \(|Q_{22}|\), as a function of GW frequency for the GR-median source parameters of GW230627\_015337. For each curve, one coupling is set to unit magnitude, \(|\beta_k|=1\), while the other two couplings are set to zero. The \(\beta_2\) curve contains both its 5PN and 6PN contributions. The vertical dotted line denotes the GR \((2,2)\)-mode peak frequency and, for \(F=1\), the center of the Fermi--Dirac taper for this mode. Since the window acts on the second derivative of the phase correction, the integrated correction does not in general vanish above the taper frequency.}
\end{figure*}

Figure~\ref{fig:dephasing} shows the magnitude of the final tapered phase correction,
\(|Q_{22}(f)|\), evaluated using the GR-median source parameters of GW230627\_015337. For each curve, one EFT coupling is set to unit magnitude while the other two are set to zero. The \(\beta_2\) result includes the linked 5PN and 6PN contributions. The vertical line indicates the \((2,2)\)-mode peak frequency, which coincides with the center of the Fermi--Dirac window for \(F=1\). Since the window is applied to the second derivative of the phase rather than directly to the phase itself, \(Q_{22}\) does not in general vanish above the taper frequency.

\section{Fisher estimation and event selection}
\label{sec:3}
\renewcommand{\theequation}{3.\arabic{equation}}
\setcounter{equation}{0}

A full Bayesian analysis of a large number of BBH events is computationally expensive. We therefore use a Fisher-matrix analysis as a first step to estimate the expected constraining power of different events and to identify the most promising target for the subsequent Bayesian analysis. The Fisher calculation is performed using the EFT-modified waveform constructed in Sec.~\ref{sec:fti} and under a common set of assumptions for all events, which allows a consistent comparison across the sample. We emphasize that the Fisher analysis is used here for forecasting and event ranking, rather than as a substitute for the full Bayesian parameter estimation. In this section, we first introduce the Fisher-matrix formalism, then describe the source-selection criteria, and finally present the forecast constraints for the selected events.

\subsection{Fisher-matrix formalism}

For an event \(e\), let \(h_d(\vec{\theta})\) denote the waveform response in detector \(d\) for a parameter vector \(\vec{\theta}\). We define the network noise-weighted inner product as
\bqn
(a|b)
\equiv
4\,{\rm Re}
\sum_{d\in{\cal D}_e}
\int_{{\cal B}_{e,d}}
\frac{
\tilde a_d^*(f)\tilde b_d(f)
}{
S_{n,d}^{(e)}(f)
}\,df ,
\eqn
where \({\cal D}_e\) denotes the set of detectors used for event \(e\),
\(S_{n,d}^{(e)}(f)\) is the corresponding one-sided noise power spectral density (PSD), and
\({\cal B}_{e,d}\) denotes the frequency region retained for that detector. The Fisher information matrix is then
\bqn
\Gamma_{ij}
\equiv
\left(
\frac{\partial h}{\partial\theta_i}
\middle|
\frac{\partial h}{\partial\theta_j}
\right),
\eqn
where \(h\) represents the network detector response. In the local Gaussian approximation, and in the absence of informative priors, the marginalized uncertainty on a parameter \(\theta_i\) is
\bqn
\sigma_i
=
\sqrt{
\left(\Gamma^{-1}\right)_{ii}
},
\eqn
provided that the Fisher matrix is nonsingular
\cite{Cutler:1994ys,Vallisneri:2007ev}.

We perform a separate Fisher analysis for each EFT coupling \(\beta_k\), varying one coupling at a time while fixing the other two to zero. The GR parameter vector contains eleven parameters,
\bqn
\vec\theta_{\rm GR}
=
(m_1,m_2,\chi_{1z},\chi_{2z},
d_L,\iota,\alpha,\delta,\psi,t_c,\phi_c),
\eqn
where \(m_1\) and \(m_2\) are the detector-frame component masses,
\(\chi_{1z}\) and \(\chi_{2z}\) are the spin components aligned with the orbital angular momentum,
\(d_L\) is the luminosity distance,
\(\iota\) is the inclination angle,
\((\alpha,\delta)\) specify the sky position,
\(\psi\) is the polarization angle,
and \(t_c\) and \(\phi_c\) are the coalescence time and phase, respectively.
Including one EFT coupling gives a 12-dimensional parameter space. The uncertainty on \(\beta_k\) is obtained after marginalizing over all eleven GR parameters.

For each event, the Fisher matrix is evaluated at a fiducial GR parameter point constructed from the marginal posterior medians reported in the corresponding
\texttt{IMRPhenomXPHM} or
\texttt{IMRPhenomXPHM-SpinTaylor}
parameter-estimation results
\cite{Pratten:2020ceb,Colleoni:2024knd}.\footnote{The PE releases used here do not provide independent \texttt{SEOBNRv5HM\_ROM} posterior samples for the event sample. Independent posteriors from the related precessing model \texttt{SEOBNRv5PHM} are available for only part of the sample. We therefore use the available \texttt{IMRPhenomXPHM} or \texttt{IMRPhenomXPHM-SpinTaylor} posteriors to construct the fiducial points across the sample using a common prescription.}
For each posterior spin sample, we first project the spin onto the orbital angular momentum and then take the median of the resulting aligned-spin component. The reported inclination angle \(\iota\) is used directly. The resulting fiducial point is therefore assembled from separate one-dimensional posterior medians and should not be interpreted as a joint posterior sample. It was also obtained using a waveform model different from the \texttt{SEOBNRv5HM\_ROM} waveform used in our Fisher calculation. We evaluate the waveform directly at this fiducial point without rescaling its amplitude to reproduce the catalog SNR. The catalog SNR is used only in the source-selection procedure described below and for the color scale in Fig.~\ref{fig:mass}.

For numerical differentiation, we use logarithmic coordinates for the component masses and luminosity distance. The waveform-shape derivatives are evaluated using fourth-order central finite differences with a step size of \(10^{-4}\) in the corresponding numerical coordinates. The derivatives associated with the detector response are evaluated separately, while those with respect to luminosity distance and coalescence phase are calculated analytically. The derivative with respect to \(\beta_k\) is obtained from the FTI-modified waveform described in Sec.~\ref{sec:fti}. The same detector responses and seven waveform modes used in the Bayesian analysis are included in the Fisher calculation. Since the masses and luminosity distance are nuisance parameters, the use of logarithmic coordinates for these quantities does not change the marginalized uncertainty on \(\beta_k\).

The inner products are evaluated on a frequency grid from \(20\) to
\(4095.75\,{\rm Hz}\), with a spacing of \(0.125\,{\rm Hz}\). For each detector, the PSD is linearly interpolated only within its supplied frequency range, and the inner product is evaluated over the intersection between this range and the analysis grid. This defines the detector-dependent frequency support used in the Fisher forecast. It does not reproduce all detector-specific frequency cuts adopted in the original parameter-estimation analyses; in particular, the presence of a PSD value at a given frequency does not necessarily imply that the corresponding strain data were included in those analyses.

The \texttt{SEOBNRv5HM\_ROM} modes have mode-dependent hard frequency terminations. Near these boundaries, finite-difference derivatives can be affected by the motion of the waveform cutoff rather than by the smooth dependence of the waveform on the physical parameters. To avoid this numerical effect, we exclude frequency bins within \(2\,{\rm Hz}\) of each mode termination. The resulting frequency mask is then held fixed when evaluating all waveform derivatives and Fisher inner products.

To marginalize over the GR parameters in a numerically stable way, we work directly with the noise-weighted waveform derivatives. We first normalize the eleven GR derivative vectors and use singular-value decomposition (SVD) to determine the subspace spanned by them. Singular values smaller than \(10^{-10}\) times the largest singular value are discarded. Let \(P_{\rm GR}\) denote the orthogonal projector onto the retained GR derivative subspace, and let \(g_k\) denote the noise-weighted derivative with respect to \(\beta_k\). The marginalized Fisher uncertainty can then be written as
\bqn
\sigma(\beta_k)
=
\left\|
(I-P_{\rm GR})g_k
\right\|^{-1}.
\label{eq:fisher_sigma_beta}
\eqn
This expression is equivalent to marginalizing over the GR nuisance parameters in the Fisher matrix, while providing better numerical stability when some derivative directions are nearly degenerate. We retain only events for which the GR derivative block is numerically full rank under the adopted SVD criterion. All 92 events in the selected sample satisfy this requirement.

Finally, we convert the Fisher uncertainty on each dimensionless coupling into a forecast constraint on the corresponding characteristic length,
\bqn
\ell_{90}(\beta_k)
=
r_g
\left[
1.645\,\sigma(\beta_k)
\right]^{1/p_k},
\label{eq:fisher_length}
\eqn
where \(p_k=4\) for \(k=1,2\) and \(p_k=6\) for \(k=3\), following Eq.~(\ref{eq:ell_def}). Here
\(r_g=GM_{\rm src}/c^2\),
with \(M_{\rm src}\) fixed at its fiducial value for each event. The source-frame mass entering this conversion is distinct from the detector-frame masses used in waveform generation. For a zero-mean Gaussian distribution of \(\beta_k\), the factor \(1.645\) gives the 90\% quantile of \(|\beta_k|/\sigma(\beta_k)\). The resulting \(\ell_{90}\) should therefore be interpreted as a local Gaussian forecast rather than as a calibrated Bayesian credible limit or frequentist confidence limit. This distinction becomes particularly important when the Gaussian approximation is poor, as can occur for signals with relatively low waveform SNR \cite{Vallisneri:2007ev}.

\subsection{Event selection}

After combining GWTC-2.1 v2, GWTC-3 v3, GWTC-4.1, and GWTC-5.0, removing duplicate entries, and excluding systems containing neutron stars, we obtain a parent sample of 273 BBH events
\cite{PE:GWTC21,PE:GWTC3,PE:GWTC41,PE:GWTC5}.
We further require the selected parameter-estimation (PE) result for each event to provide PSDs for at least two detectors, so that the Fisher forecast can be performed with a multi-detector noise model. This requirement removes nine events and leaves 264 candidates.

For each candidate, we use the GR parameter medians and detector PSDs from the same PE result group. The event parameters, catalog SNR, false-alarm rate (FAR), detector network, and corresponding PE information are recorded together so that the source parameters and noise spectra used in the Fisher calculation can be traced consistently to the same PE analysis.

Since the EFT corrections considered here affect the inspiral phase, we select events with sufficiently strong and long inspiral signals. We characterize the end of the inspiral interval by the frequency \(f_{\rm c}\), defined through
\bqn
\frac{G M_{\rm det} f_{\rm c}}{c^3}=0.018,
\label{eq:fc}
\eqn
where \(M_{\rm det}\) is the detector-frame total mass. This choice follows the dimensionless-frequency boundary \(Mf=0.018\) used to separate the inspiral and merger regimes in the \texttt{IMRPhenomD} phase model
\cite{Husa:2015iqa,Khan:2015jqa} and was also adopted in Ref.~\cite{Liu:2024atc}. We use \(f_{\rm c}\) only to estimate the amount of inspiral information available for event selection. It should not be interpreted as an EFT-validity cutoff and is independent of the FTI taper introduced in Sec.~\ref{sec:fti}.

We estimate the inspiral contribution to the network SNR by computing the fraction of the squared SNR accumulated below \(f_{\rm c}\) and normalizing it to the catalog-reported network SNR,
\bqn
\rho_{\rm insp}
=
\rho_{\rm cat}
\left[
\frac{
\displaystyle
\sum_d
\int_{20\,{\rm Hz}}^{f_{\rm c}}
\frac{|\tilde h_d(f)|^2}{S_{n,d}(f)}\,df
}{
\displaystyle
\sum_d
\int_{20\,{\rm Hz}}^{f_{\rm max}}
\frac{|\tilde h_d(f)|^2}{S_{n,d}(f)}\,df
}
\right]^{1/2}.
\label{eq:rhoinsp}
\eqn
The integrals are evaluated using the event-specific detector PSDs and only over their available frequency support. The waveform is evaluated at the PE-median GR parameters. Since only the relative SNR contribution is used in Eq.~(\ref{eq:rhoinsp}), an overall amplitude normalization cancels in the ratio. We emphasize that \(\rho_{\rm insp}\) is a catalog-normalized quantity used only for event selection and is not the optimal SNR of the fiducial waveform used in the Fisher calculation.

As a second measure of the available inspiral information, we estimate the number of GW cycles between \(f_{\rm min}=20\,{\rm Hz}\) and \(f_{\rm c}\) at leading PN order,
\bqn
N_{\rm c}
\simeq
\frac{1}{32\pi^{8/3}}
\left(
\frac{G{\cal M}_{\rm det}}{c^3}
\right)^{-5/3}
\left(
f_{\rm min}^{-5/3}
-f_{\rm c}^{-5/3}
\right),
\label{eq:Ncycle}
\eqn
where
\({\cal M}_{\rm det}=M_{\rm det}\eta^{3/5}\)
is the detector-frame chirp mass. We set
\(N_{\rm c}=0\)
when \(f_{\rm c}\leq f_{\rm min}\).
This leading-order estimate is used only as an event-selection criterion
\cite{Blanchet:2013haa}.

We retain events satisfying
\bqn
\rho_{\rm insp}>6,
\qquad
N_{\rm c}>5,
\qquad
{\rm FAR}<10^{-3}\,{\rm yr}^{-1}.
\label{eq:cut}
\eqn
Together with the requirement of PSD availability for at least two detectors, these cuts select significant events with appreciable inspiral information. We adopt a relatively mild threshold on \(\rho_{\rm insp}\) because the characteristic length depends only on the fourth or sixth root of the coupling uncertainty. The Fisher results for events near the threshold should nevertheless be regarded only as local forecasts, rather than as precise parameter-estimation bounds. Applying these criteria leaves 92 BBH events. All of them also satisfy the numerical full-rank condition required in the Fisher calculation. 
 %The complete parent sample, selection quantities, selection flags, and PE information are provided in the accompanying data table.

\begin{table*}[tbp]
\caption{\label{selection_table} The twenty tightest Fisher forecasts with eleven GR nuisance parameters among the 92 selected multi-detector events, ordered by $\ell_{90}(\beta_2)$. $M_{\rm det}$ and $M_{\rm src}$ are evaluated at the PE-median point; $\rho_{\rm cat}$ is the catalog network SNR and $\rho_{\rm insp}$ is the catalog-normalized inspiral SNR computed with the event's PE Release PSDs. The final three columns are one-sided 90\% Fisher forecasts for the characteristic lengths, not Bayesian bounds.}
\begin{ruledtabular}
\normalfont\normalsize
\setlength{\tabcolsep}{1.5pt}
\begin{tabular}{lccccccccccc}
Event & $M_{\rm det}$ & $M_{\rm src}$ & $\eta$ & $f_{\rm c}$ & $\rho_{\rm cat}$ & $\rho_{\rm insp}$ & $N_{\rm c}$ & FAR & \multicolumn{3}{c}{$\ell_{90}$ [km]} \\
~ & [$M_\odot$] & [$M_\odot$] & ~ & [Hz] & ~ & ~ & ~ & [$\mathrm{yr}^{-1}$] & $\beta_1$ & $\beta_2$ & $\beta_3$ \\
\hline
GW230627\_015337 & 15.08 & 14.14 & 0.241 & 242.4 & 28.7 & 26.6 & 312.4 & $1\times10^{-5}$ & 16.9 & 11.1 & 17.9 \\
GW240925\_005809 & 17.07 & 15.85 & 0.247 & 214.1 & 31.2 & 28.8 & 246.7 & $1\times10^{-5}$ & 17.4 & 11.4 & 19.1 \\
GW240413\_022019 & 15.02 & 13.42 & 0.242 & 243.3 & 18.1 & 17.0 & 312.0 & $1\times10^{-5}$ & 18.5 & 12.2 & 18.7 \\
GW241109\_115924 & 13.91 & 12.37 & 0.243 & 262.7 & 11.9 & 11.3 & 354.8 & $1.2\times10^{-4}$ & 21.0 & 13.8 & 19.9 \\
GW240910\_103535 & 19.77 & 17.19 & 0.232 & 184.8 & 17.8 & 16.1 & 204.2 & $1\times10^{-5}$ & 21.9 & 14.5 & 22.2 \\
GW190924\_021846 & 15.81 & 14.20 & 0.224 & 231.2 & 12.0 & 11.5 & 310.3 & $1\times10^{-5}$ & 22.6 & 15.0 & 21.3 \\
GW231224\_024321 & 19.66 & 16.55 & 0.246 & 185.8 & 14.0 & 12.7 & 194.1 & $1\times10^{-5}$ & 23.5 & 15.5 & 23.4 \\
GW241011\_233834 & 26.67 & 25.49 & 0.179 & 137.0 & 35.8 & 29.3 & 158.0 & $1\times10^{-5}$ & 23.4 & 15.6 & 24.7 \\
GW241102\_124058 & 20.48 & 18.91 & 0.244 & 178.5 & 21.7 & 19.3 & 182.7 & $1\times10^{-5}$ & 23.8 & 15.7 & 24.6 \\
GW250119\_190238 & 23.06 & 21.61 & 0.249 & 158.5 & 21.3 & 18.5 & 146.4 & $1\times10^{-5}$ & 24.0 & 15.8 & 25.8 \\
GW240915\_001357 & 21.99 & 18.57 & 0.243 & 166.2 & 16.4 & 14.5 & 162.7 & $1\times10^{-5}$ & 24.6 & 16.2 & 24.8 \\
GW230630\_234532 & 20.15 & 16.62 & 0.240 & 181.4 & 9.9 & 8.7 & 191.0 & $4.2\times10^{-4}$ & 24.6 & 16.2 & 24.1 \\
GW241225\_042553 & 23.21 & 20.43 & 0.239 & 157.4 & 16.5 & 14.4 & 150.8 & $1\times10^{-5}$ & 24.6 & 16.3 & 25.5 \\
GW241231\_054133 & 22.99 & 19.44 & 0.231 & 158.9 & 17.4 & 15.2 & 158.2 & $1\times10^{-5}$ & 25.1 & 16.6 & 25.3 \\
GW240629\_145256 & 23.34 & 18.74 & 0.243 & 156.5 & 12.7 & 11.1 & 146.9 & $1\times10^{-5}$ & 25.8 & 17.1 & 25.9 \\
GW240922\_142106 & 23.63 & 19.08 & 0.241 & 154.7 & 12.5 & 10.7 & 145.0 & $1\times10^{-5}$ & 25.8 & 17.1 & 25.8 \\
GW191216\_213338 & 21.44 & 19.99 & 0.232 & 170.5 & 18.6 & 16.7 & 177.5 & $1\times10^{-5}$ & 26.1 & 17.3 & 26.2 \\
GW170608 & 19.85 & 18.53 & 0.243 & 184.1 & 15.3 & 14.3 & 193.8 & $1\times10^{-7}$ & 26.3 & 17.4 & 26.3 \\
GW241114\_235258 & 21.89 & 18.86 & 0.241 & 166.9 & 12.4 & 11.0 & 165.1 & $1\times10^{-5}$ & 26.5 & 17.5 & 26.3 \\
GW241130\_110422 & 21.90 & 17.49 & 0.233 & 166.9 & 10.3 & 9.2 & 170.6 & $5.5\times10^{-4}$ & 26.9 & 17.8 & 25.7 \\
\end{tabular}
\end{ruledtabular}
\end{table*}

\subsection{Forecast constraints for the selected events}
\label{sec:3c}

\begin{figure*}[tbp]
\centering
\includegraphics[width=\textwidth]{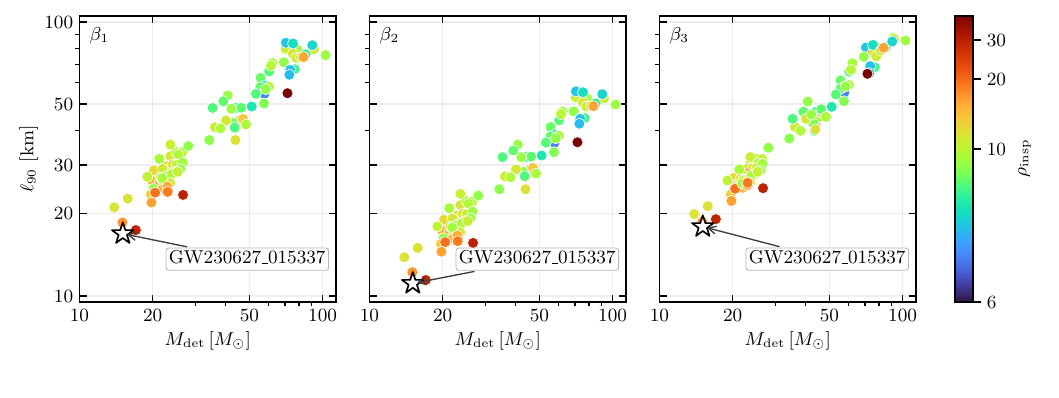}
\caption{\label{fig:mass}
Fisher forecast constraints on the characteristic lengths associated with \(\beta_1\) (left), \(\beta_2\) (middle), and \(\beta_3\) (right) for the 92 selected BBH events, shown as a function of detector-frame total mass. The color scale represents the catalog-normalized inspiral SNR defined in Eq.~(\ref{eq:rhoinsp}). The stars indicate GW230627\_015337, which gives the tightest forecast for all three couplings in the selected sample.}
\end{figure*}

Figure~\ref{fig:mass} shows the Fisher forecasts for the 92 selected events. The forecast characteristic lengths are plotted against the detector-frame total mass, with the color indicating the catalog-normalized inspiral SNR. The conversion from the dimensionless couplings to the characteristic lengths is performed using the corresponding source-frame masses according to Eq.~(\ref{eq:ell_def}). Table~\ref{selection_table} lists the twenty events with the tightest forecasts, ordered by \(\ell_{90}(\beta_2)\). The same waveform model and Fisher procedure are applied to all events, while the source parameters and detector PSDs are taken from the corresponding PE analyses.

As seen in Fig.~\ref{fig:mass}, the constraining power is not determined by the inspiral SNR alone. The characteristic length depends explicitly on the source mass, while the uncertainty on the EFT coupling enters only through a fourth or sixth root. The source masses also determine the duration and frequency range of the inspiral signal and affect correlations between the EFT coupling and the GR parameters. In addition, the detector responses and noise spectra determine the frequency-dependent weighting of the signal. These effects are included simultaneously in the Fisher calculation. Consequently, events with similar inspiral SNRs can give noticeably different constraints, and a larger SNR does not necessarily imply a tighter bound on the characteristic length.

Among the 92 selected events, GW230627\_015337 provides the tightest forecast for all three EFT couplings, with
\(\ell_{90}=16.9\), \(11.1\), and \(17.9\,{\rm km}\)
for \(\beta_1\), \(\beta_2\), and \(\beta_3\), respectively. For comparison, GW240925\_005809 gives the second-tightest constraint on \(\beta_2\), with
\(\ell_{90}(\beta_2)=11.4\,{\rm km}\), while GW170608 gives
\(\ell_{90}(\beta_2)=17.4\,{\rm km}\). The latter event was used in the previous cubic-EFT analysis of Ref.~\cite{Liu:2024atc}. Based on this ranking, we select GW230627\_015337 for the full Bayesian analysis presented in the next section. We emphasize that this ranking refers only to the 92 events passing the selection criteria of Sec.~\ref{sec:3}; it should not be interpreted as a ranking of all 273 BBH events in the parent sample.

\section{Bayesian inference and analysis setup}
\label{sec:4}
\renewcommand{\theequation}{4.\arabic{equation}}
\setcounter{equation}{0}

Having identified GW230627\_015337 as the most promising event from the Fisher analysis, we now perform a full Bayesian parameter-estimation analysis using its strain data. Unlike the Fisher forecast, which relies on a local Gaussian approximation around a fiducial point, Bayesian inference samples the posterior distribution over the full parameter space under specified priors.

For a waveform model \(H\) and data \(d\), Bayes' theorem gives
\bqn
p(\vec{\theta}\mid d,H)
=
\frac{
{\cal L}(d\mid\vec{\theta},H)
\pi(\vec{\theta}\mid H)
}{
{\cal Z}(d\mid H)
},
\eqn
where
\({\cal L}(d\mid\vec{\theta},H)\)
is the likelihood,
\(\pi(\vec{\theta}\mid H)\)
is the prior, and
\({\cal Z}(d\mid H)\)
is the Bayesian evidence. For two models \(H_A\) and \(H_B\), we define the Bayes factor as
\bqn
{\cal B}^{A}_{B}
=
\frac{{\cal Z}(d\mid H_A)}
     {{\cal Z}(d\mid H_B)}.
\eqn
Values
\({\cal B}^{A}_{B}>1\)
indicate that the data favor \(H_A\) over \(H_B\) for the adopted priors.

The parameter space contains the standard BBH parameters used in the GR analysis: the component masses, aligned-spin components, luminosity distance, inclination angle, coalescence time and phase, sky position, and polarization angle. For each EFT model, we vary one coupling \(\beta_k\) at a time while fixing the other two to zero. The corresponding GR model is recovered at \(\beta_k=0\).

We perform the parameter estimation using the open-source package
\texttt{bilby} \cite{Ashton:2018jfp}
together with the nested-sampling algorithm
\texttt{nessai} \cite{Williams:2021qyt}.
The waveform model is the EFT-modified
\texttt{SEOBNRv5HM\_ROM}
waveform described in Sec.~\ref{sec:fti}, implemented using
\texttt{LALSuite}
\cite{LALSuite,Pompili:2023tna}.
The main settings of the analysis are summarized in Table~\ref{tab:settings}.

\begin{table}
\caption{\label{tab:settings}
Main settings used in the Bayesian analysis of GW230627\_015337.}
\begin{ruledtabular}
\begin{tabular}{ll}
Quantity & Setting \\
\hline
Event & GW230627\_015337 \\
Detectors & H1, L1 \\
GPS trigger time & 1371866035.8 \\
Segment duration & 16 s \\
Sampling rate & 4096 Hz \\
Frequency range & 20--1024 Hz \\
Reference frequency & 20 Hz \\
Strain data & GWOSC, O4a, 4 kHz R1 \\
PSD & PE Release H1/L1 PSDs \\
Baseline waveform & \texttt{SEOBNRv5HM\_ROM} \\
Harmonics & (2,2), (2,1), (3,3), (3,2), \\
~ & (4,4), (4,3), (5,5) \\
EFT corrections & 5PN, 6PN, 8PN phase corrections \\
FTI window & \(F=1.0,\;N_{\rm cyc}=1\) \\
Prior on \(\beta_{1,2}\) & \(U(-20,20)\) \\
Prior on \(\beta_3\) & \(U(-20/3,20/3)\) \\
Marginalization & time and distance \\
Sampler & \texttt{nessai} \\
Live points & 1000 \\
Stopping criterion & 0.1 \\
Parallel pool & 64 \\
\end{tabular}
\end{ruledtabular}
\end{table}

We perform four analyses using the same strain data, likelihood, waveform settings, and priors on the GR parameters: one GR analysis and three single-coupling EFT analyses corresponding to \(\beta_1\), \(\beta_2\), and \(\beta_3\). In each EFT analysis, the other two couplings are fixed to zero. Each EFT model therefore contains GR as the zero-coupling limit, which allows the corresponding evidences to be compared directly under the specified priors.

We analyze a \(16\,{\rm s}\) strain segment extending from \(14\,{\rm s}\) before to \(2\,{\rm s}\) after the trigger time, and use the frequency range \(20\)--\(1024\,{\rm Hz}\). The reference frequency is set to
\(f_{\rm ref}=20\,{\rm Hz}\).
For the \((2,2)\) mode, the phase correction is therefore anchored at the lower analysis frequency. For the other modes, the corresponding reference frequency is
\bqn
f_{{\rm ref},m}
=
\frac{m}{2}f_{\rm ref},
\eqn
while the second integration condition is imposed at
\bqn
f_{{\rm a},m}
=
\frac{m}{2}f_{22}^{\rm peak},
\eqn
as described in Sec.~\ref{sec:fti}. For the choice \(F=1\), this frequency also coincides with the center of the FTI taper for each mode.

The coupling priors in the convention used throughout this paper are
\bqn
\beta_1,\beta_2
&\sim&
U(-20,20),
\nb\\
\beta_3
&\sim&
U(-20/3,20/3).
\eqn
The different numerical range for \(\beta_3\) originates from the convention used in the archived analysis code,
\(\beta_3^{\rm code}=3\beta_3\).
Thus, a uniform prior
\(\beta_3^{\rm code}\sim U(-20,20)\)
corresponds exactly to
\(\beta_3\sim U(-20/3,20/3)\)
in the convention adopted here. This is only a change of parameter convention and does not represent an additional physical restriction on the posterior. The posterior support is well contained within the adopted prior ranges.

\section{Results}\label{sec:5}
\renewcommand{\theequation}{5.\arabic{equation}} \setcounter{equation}{0}

\begin{figure*}[tbp]
\centering
\includegraphics[width=\textwidth]{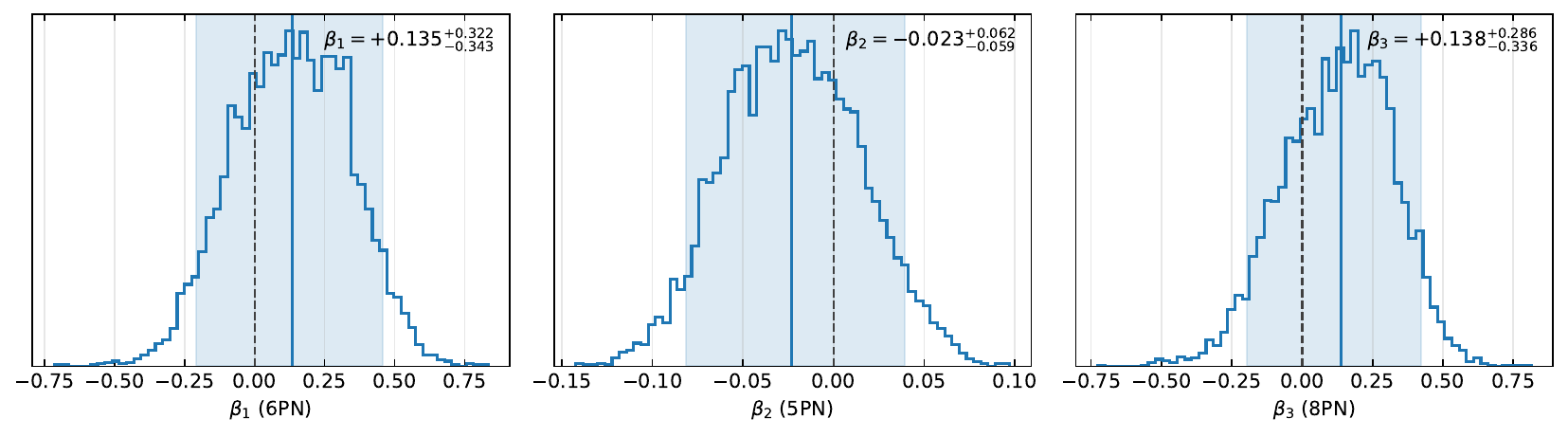}
\caption{\label{fig:post}
One-dimensional posterior distributions of \(\beta_1\), \(\beta_2\), and \(\beta_3\) for GW230627\_015337. The dashed vertical lines indicate the GR value \(\beta_k=0\), while the solid lines and shaded regions show the posterior medians and central 90\% credible intervals, respectively.}
\end{figure*}

The posterior distributions of the three EFT couplings for GW230627\_015337 are shown in Fig.~\ref{fig:post}, and the corresponding numerical results are summarized in Table~\ref{tab:results}.

\begin{table}[tbp]
\caption{\label{tab:results}
Bayesian constraints on the EFT couplings and characteristic length scales for GW230627\_015337. The second column gives the posterior median and central 90\% credible interval, while the third column gives the one-sided 90\% upper limit on the corresponding characteristic length.}
\begin{ruledtabular}
\begin{tabular}{lccc}
Coupling &
90\% credible interval &
\(\ell_{90}\) [km] &
\(\ln\mathcal B^{\rm EFT}_{\rm GR}\) \\
\hline
\(\beta_1\) &
\(0.135^{+0.322}_{-0.343}\) &
\(<17.1\) &
\(-4.10\) \\
\(\beta_2\) &
\(-0.023^{+0.062}_{-0.059}\) &
\(<11.1\) &
\(-5.49\) \\
\(\beta_3\) &
\(0.138^{+0.286}_{-0.336}\) &
\(<18.3\) &
\(-2.51\) \\
\end{tabular}
\end{ruledtabular}
\end{table}

\subsection{Constraints on the EFT couplings}

The 90\% credible intervals of all three EFT couplings contain zero. We therefore find no evidence for a deviation from GR. The constraints are
\bqn
\beta_1
&=&
0.135^{+0.322}_{-0.343},
\qquad
\beta_2
=
-0.023^{+0.062}_{-0.059},
\nb\\
\beta_3
&=&
0.138^{+0.286}_{-0.336},
\eqn
where the uncertainties denote the central 90\% credible intervals.

Among the three couplings, \(\beta_2\) is most tightly constrained. This is consistent with the fact that \(\beta_2\) contributes already at 5PN order and also produces a linked 6PN correction, whereas \(\beta_1\) first enters at 6PN order and \(\beta_3\) at 8PN order.

Using Eq.~(\ref{eq:ell_def}), we convert the joint posterior samples of the EFT coupling and source-frame mass into the corresponding characteristic lengths. The resulting one-sided 90\% upper limits are
\bqn
\ell_1 < 17.1\,{\rm km},
\qquad
\ell_2 < 11.1\,{\rm km},
\qquad
\ell_3 < 18.3\,{\rm km}.
\label{eq:l90final}
\eqn
The strongest constraint is therefore obtained for the coupling \(\beta_2\).

We quote upper limits on the characteristic lengths because the sampling priors are uniform in the signed couplings \(\beta_k\), rather than in \(\ell_k\). Since
\(\ell_k\propto|\beta_k|^{1/p_k}\),
the induced prior on the characteristic length is not uniform, and its posterior need not peak at zero even when the coupling posterior is consistent with GR.

\subsection{Bayesian model comparison}

We compare each single-coupling EFT model with GR using
\(\mathcal B^{\rm EFT}_{\rm GR}
=\mathcal Z_{\rm EFT}/\mathcal Z_{\rm GR}\).
The corresponding log Bayes factors are
\[
\ln\mathcal B^{\rm EFT}_{\rm GR}
=
-4.10,\;-5.49,\;-2.51
\]
for the \(\beta_1\), \(\beta_2\), and \(\beta_3\) models, respectively. All three values favor GR over the corresponding EFT extension under the adopted priors. The strongest preference for GR is obtained for the \(\beta_2\) model.

The numerical values of the Bayes factors depend on the adopted prior ranges, as expected in Bayesian model comparison. The coupling posteriors are well contained within the priors used in our analysis.

\subsection{Comparison with the Fisher forecasts}

For GW230627\_015337, the Fisher forecasts for the characteristic lengths are
\(16.9\), \(11.1\), and \(17.9\,{\rm km}\)
for \(\beta_1\), \(\beta_2\), and \(\beta_3\), respectively. These values are close to the Bayesian upper limits of
\(17.1\), \(11.1\), and \(18.3\,{\rm km}\).

The Fisher and Bayesian analyses are based on different statistical assumptions: the Fisher calculation uses a local Gaussian approximation, while the Bayesian analysis samples the full posterior under explicit priors. Nevertheless, the close agreement of the resulting characteristic-length constraints shows that the Fisher analysis provides a useful estimate of the relative constraining power of different events. This supports its use for the event selection carried out in Sec.~\ref{sec:3c}.

\subsection{Comparison with previous constraints}
\label{sec:compare}

We compare our constraints with previous inspiral and ringdown studies of higher-curvature EFTs in Table~\ref{tab:compare}. Such comparisons should be interpreted with some care. The analyses use different events, waveform models, statistical methods, priors, and prescriptions for the regime of validity of the EFT. In addition, the cubic theories considered in different studies are not always parametrized by the same operator basis. We therefore use the previous results mainly to compare the characteristic scales that can be probed, rather than to define precise improvement factors.

For the quartic sector, Ref.~\cite{Sennett:2019bpc} analyzed GW151226 and GW170608 and found that EFT length scales around
\(d_\Lambda\sim150\,{\rm km}\)
are strongly disfavored by Bayesian model selection. This result is not a 90\% upper limit and therefore cannot be directly compared with our bound
\[
\ell_3<18.3\,{\rm km}.
\]
Nevertheless, the smaller scale reached in the present analysis indicates the increased sensitivity provided by the newer event and waveform treatment. 
%We also note an important difference in the treatment of EFT validity. Ref.~\cite{Sennett:2019bpc} imposed an upper frequency related to the EFT cutoff, whereas the FTI taper used here controls the transition of the phase correction to the GR merger-ringdown waveform. The FTI taper should not itself be interpreted as an EFT-validity condition.
Ringdown constraints provide a complementary comparison. Combining GW150914 and GW200129, Ref.~\cite{Silva:2022srr} obtained the 90\% upper limits
\(\ell_{\rm qEFT}<51.3\,{\rm km}\)
and
\(\ell_{\rm cEFT}<38.2\,{\rm km}\)
for the quartic and cubic EFT models considered there. The quartic result can be compared qualitatively with our
\(\ell_3<18.3\,{\rm km}\), although the ringdown analysis uses different waveform information and its quartic bound is subject to the EFT-validity qualification discussed in Ref.~\cite{Silva:2022srr}. The cubic comparison is less direct. The cubic ringdown model involves a particular choice of parity-even and parity-odd couplings, whereas our analysis treats the two parity-even couplings associated with \(I_1\) and \(\mathcal G_3\) separately. The value
\(38.2\,{\rm km}\)
should therefore be regarded only as a reference scale and not as a direct bound on either \(\beta_1\) or \(\beta_2\).

The cubic sector has also been studied using GW170608 in Ref.~\cite{Liu:2024atc}, where an approximate inspiral-merger-ringdown waveform was constructed for a parity-preserving cubic EFT and analyzed in a full Bayesian framework. Because the waveform model, coupling parametrization, and treatment of EFT corrections differ from those adopted here, we do not make a direct numerical comparison between the corresponding coupling bounds. A more direct comparison can instead be made within our own Fisher framework. For GW170608, we obtain
\(\ell_{90}(\beta_1)=26.3\,{\rm km}\)
and
\(\ell_{90}(\beta_2)=17.4\,{\rm km}\),
compared with
\(16.9\,{\rm km}\)
and
\(11.1\,{\rm km}\)
for GW230627\_015337. Thus, under the same waveform and statistical assumptions, the event selected in this work is expected to provide substantially tighter constraints than GW170608.

Overall, the present analysis probes higher-curvature length scales at the level of approximately \(10\)--\(20\,{\rm km}\). The comparison with previous studies shows the improvement in the accessible scale, but the numerical differences should not be interpreted as precise improvement factors unless the waveform model, operator normalization, priors, and EFT-validity conditions are treated consistently.

\begin{table}[tbp]
\caption{\label{tab:compare}
Comparison with representative previous constraints on higher-curvature EFT length scales. The quartic inspiral value from Ref.~\cite{Sennett:2019bpc} denotes a scale disfavored by Bayesian model selection rather than a 90\% upper limit. The cubic ringdown result of Ref.~\cite{Silva:2022srr} corresponds to a different cubic model and is included only as a reference scale.}
\begin{ruledtabular}
\begin{tabular}{lccc}
Sector & This work & Previous inspiral & Ringdown \\
\hline
Cubic \(\beta_1\) &
\(<17.1\) km &
Ref.~\cite{Liu:2024atc} &
\(<38.2\) km$^{a}$ \\
Cubic \(\beta_2\) &
\(<11.1\) km &
Ref.~\cite{Liu:2024atc} &
\(<38.2\) km$^{a}$ \\
Quartic \(\beta_3\) &
\(<18.3\) km &
\(\sim150\) km$^{b}$ &
\(<51.3\) km \\
\end{tabular}
\end{ruledtabular}

\begin{flushleft}
\footnotesize
$^{a}$ Different cubic coupling model; not a direct bound on \(\beta_1\) or \(\beta_2\).\\
$^{b}$ Characteristic scale disfavored by Bayesian model selection in Ref.~\cite{Sennett:2019bpc}.
\end{flushleft}
\end{table}

\section{Conclusions}
\label{sec:6}
\renewcommand{\theequation}{6.\arabic{equation}}
\setcounter{equation}{0}

In this work, we have studied higher-curvature EFT extensions of GR using the inspiral signals of binary black holes. We considered two parity-even cubic-curvature couplings and one quartic-curvature coupling, whose leading phase corrections enter at 5PN, 6PN, and 8PN order. These corrections were incorporated into the individual multipoles of the \texttt{SEOBNRv5HM\_ROM} waveform using the FTI framework, which allows the inspiral modifications to be combined with higher-order modes and the full GR inspiral-merger-ringdown waveform.

We first used a Fisher-matrix analysis to identify the events with the strongest expected constraints. Starting from a parent sample of 273 BBH events, we selected 92 systems with sufficient inspiral information and suitable detector PSDs. The Fisher calculation includes event-specific source parameters and noise spectra and marginalizes over eleven GR parameters. Among the selected events, GW230627\_015337 gives the tightest forecasts for all three EFT couplings, with characteristic-length constraints of \(16.9\), \(11.1\), and \(17.9\,{\rm km}\) for \(\beta_1\), \(\beta_2\), and \(\beta_3\), respectively. This result also shows that the most constraining event cannot be identified from the catalog SNR alone, since the source mass, inspiral duration, parameter correlations, detector responses, and noise spectra all contribute to the final constraint.

We then performed a full Bayesian analysis of GW230627\_015337. The posterior distributions of all three EFT couplings are consistent with the GR value. The corresponding 90\% upper limits on the characteristic lengths are
\[
\ell_1<17.1\,{\rm km},
\qquad
\ell_2<11.1\,{\rm km},
\qquad
\ell_3<18.3\,{\rm km}.
\]
The log Bayes factors are
\(-4.10\), \(-5.49\), and \(-2.51\)
for the \(\beta_1\), \(\beta_2\), and \(\beta_3\) models, respectively, and therefore favor GR over the corresponding single-coupling EFT extensions for the adopted priors. The Bayesian length constraints are also in close agreement with the Fisher forecasts, supporting the use of the Fisher analysis for event ranking.

Compared with previous inspiral and ringdown studies, the present analysis probes higher-curvature length scales at the level of approximately \(10\)--\(20\,{\rm km}\). These comparisons should, however, be interpreted with care because different studies use different operator normalizations, waveform models, priors, frequency ranges, and EFT-validity assumptions. In particular, the FTI taper provides a prescription for connecting the inspiral correction to the GR merger-ringdown waveform, but it should not by itself be regarded as a condition for the validity of the EFT.

Several extensions of this analysis are possible. A simultaneous analysis of multiple EFT couplings would allow correlations among the couplings to be studied directly. Combining several BBH events could further improve the constraints and reduce the dependence on any single source. It will also be useful to investigate the effects of the FTI taper, EFT corrections to the waveform amplitudes, and spin-dependent EFT contributions. These developments would provide a more complete test of higher-curvature corrections to GR with current and future GW observations.

\section*{Data and code availability}

The analysis code and accompanying data are publicly available at
\url{https://github.com/leishiv/constraints-higher-derivative-EFTGR-gw}.
The repository contains the scripts used for waveform construction, Fisher forecasts, Bayesian parameter estimation, and figure generation, together with the event-selection table and reference results.

The strain data, parameter-estimation products, detector PSDs, and waveform ROM data are obtained from the corresponding public releases
\cite{PE:GWTC21,PE:GWTC3,PE:GWTC41,PE:GWTC5,Data:ROM}.
Instructions for obtaining and using these data are provided in the repository.

\section*{Acknowledgements}

This work is supported by the National Natural Science Foundation of China under Grant Nos. 12675080, 12542053, and 12275238, the National Key Research and Development Program under Grant No. 2020YFC2201503, and the Zhejiang Provincial Natural Science Foundation of China under Grant Nos. LR21A050001 and LY20A050002, and the Fundamental Research Funds for the Provincial Universities of Zhejiang in China under Grant No. RF-A2019015. 

This research has made use of data or software obtained from the Gravitational Wave Open Science Center (gwosc.org), a service of the LIGO Scientific Collaboration, the Virgo Collaboration, and KAGRA. This material is based upon work supported by NSF's LIGO Laboratory which is a major facility fully funded by the National Science Foundation, as well as the Science and Technology Facilities Council (STFC) of the United Kingdom, the Max-Planck-Society (MPS), and the State of Niedersachsen/Germany for support of the construction of Advanced LIGO and construction and operation of the GEO600 detector. Additional support for Advanced LIGO was provided by the Australian Research Council. Virgo is funded, through the European Gravitational Observatory (EGO), by the French Centre National de Recherche Scientifique (CNRS), the Italian Istituto Nazionale di Fisica Nucleare (INFN) and the Dutch Nikhef, with contributions by institutions from Belgium, Germany, Greece, Hungary, Ireland, Japan, Monaco, Poland, Portugal, Spain. KAGRA is supported by the Ministry of Education, Culture, Sports, Science and Technology (MEXT), Japan Society for the Promotion of Science (JSPS) in Japan; National Research Foundation (NRF) and Ministry of Science and ICT (MSIT) in Korea; Academia Sinica (AS) and National Science and Technology Council (NSTC) in Taiwan, China.

OpenAI Codex and WorkBuddy were used as AI-assisted tools to support code development, data handling, numerical-analysis workflows, figure preparation, and manuscript editing. The author reviewed the code and analysis and takes responsibility for the methods and results.

The data analysis and figure preparation used \texttt{Bilby} \cite{Ashton:2018jfp}, \texttt{nessai} \cite{Williams:2021qyt}, \texttt{LALSuite} \cite{LALSuite}, \texttt{GWpy} \cite{Macleod:2021goi}, \texttt{h5py} \cite{h5py}, \texttt{NumPy} \cite{Harris:2020xlr}, \texttt{SciPy} \cite{Virtanen:2019joe}, and \texttt{Matplotlib} \cite{Hunter:2007ouj}. The implementation of the FTI phase corrections also follows the reference code in \texttt{LALSuite} and \texttt{bilby\_tgr} \cite{LALSuite,bilbyTGR}.

\end{document}